\documentclass[aps,pra,preprint,superscriptaddress,amsmath,amssymb]{revtex4-2}
\usepackage[T1]{fontenc}
\usepackage[utf8]{inputenc}
\usepackage{graphicx}
\usepackage{microtype}
\usepackage{amsmath,amssymb}
\usepackage{booktabs}
\usepackage[hidelinks]{hyperref}
\allowdisplaybreaks[2]
\usepackage{tikz}
\usetikzlibrary{arrows.meta}

\definecolor{sourceorange}{HTML}{F4AD36}
\definecolor{boxgray}{HTML}{E9EAEC}
\definecolor{ink}{HTML}{202124}
\definecolor{muted}{HTML}{5F6368}
\definecolor{rulegray}{HTML}{B5B8BC}

\tikzset{
  every node/.style={
    text=ink,
    font=\sffamily\fontsize{10}{12}\selectfont
  },
  title/.style={
    anchor=west,
    font=\sffamily\bfseries\fontsize{11}{13}\selectfont
  },
  subtitle/.style={
    anchor=west,
    text=muted,
    font=\sffamily\fontsize{9}{11}\selectfont
  },
  source/.style={
    circle,
    draw=ink,
    line width=.7pt,
    fill=sourceorange,
    minimum size=8.5mm,
    inner sep=0pt,
    font=\fontsize{11}{12}\selectfont
  },
  measure/.style={
    draw=ink,
    line width=.65pt,
    fill=boxgray,
    minimum width=14mm,
    minimum height=8mm,
    inner sep=2pt
  },
  product/.style={
    draw=ink,
    line width=.65pt,
    fill=boxgray,
    minimum width=60mm,
    minimum height=13mm,
    inner sep=4pt,
    align=center
  },
  qflow/.style={
    -{Stealth[length=2mm,width=1.4mm]},
    draw=ink,
    line width=.75pt
  },
  cflow/.style={
    qflow,
    dash pattern=on 2.5pt off 2pt
  },
  lab/.style={
    draw=rulegray,
    line width=.6pt,
    dash pattern=on 3pt off 2.2pt
  },
  note/.style={
    text=muted,
    font=\sffamily\fontsize{9}{11}\selectfont
  },
  panel/.style={
    draw=ink,
    line width=.65pt
  }
}

\newcommand{\sources}{%
  \node[source] (s1) at (1.35,1.65) {$S_1$};
  \node[source] (s2) at (3.50,1.65) {$S_2$};
  \node at (5.65,1.65) {$\cdots$};
  \node[source] (sn) at (7.80,1.65) {$S_N$};
}
\newcommand{\laboratory}{%
  \draw[lab] (.35,2.25) rectangle (8.95,6.95);
  \node[note,fill=white,inner xsep=3pt] at (4.6,2.25) {Single laboratory};
}
\newcommand{\sequentialpanel}[4]{%
  \draw[panel] (0,0) rectangle (9.25,7.75);
  \node[title] at (.3,.43) {#1};
  \node[subtitle] at (.3,.94) {Same global box, then individual measurements};
  \sources\laboratory
  \node[measure,minimum width=72mm] (global) at (4.6,3.1) {$#2$};
  \draw[qflow] (s1.south) -- (1.35,2.7);
  \draw[qflow] (s2.south) -- (3.5,2.7);
  \draw[qflow] (sn.south) -- (7.8,2.7);
  \draw[cflow] (global.east) -- (8.85,3.1)
    node[midway,above,font=\small] {$#4$};
  \node[measure,minimum width=12mm] (m1) at (1.35,5.1) {$#3^{(1)}$};
  \node[measure,minimum width=12mm] (m2) at (3.5,5.1) {$#3^{(2)}$};
  \node at (5.65,5.1) {$\cdots$};
  \node[measure,minimum width=12mm] (mn) at (7.8,5.1) {$#3^{(N)}$};
  \draw[qflow] (1.35,3.5) -- (m1.north);
  \draw[qflow] (3.5,3.5) -- (m2.north);
  \draw[qflow] (7.8,3.5) -- (mn.north);
  \draw[qflow] (m1.south) -- (1.35,6.6);
  \draw[qflow] (m2.south) -- (3.5,6.6);
  \draw[qflow] (mn.south) -- (7.8,6.6);
}

\newcommand{\ii}{\mathrm{i}}
\newcommand{\Tr}{\operatorname{Tr}}
\newcommand{\ket}[1]{\lvert #1\rangle}
\newcommand{\bra}[1]{\langle #1\rvert}
\newcommand{\Repart}{\operatorname{Re}}

\begin{document}

\title{Network contextuality beyond binary outcomes}

\author{Hao Zhang}
\affiliation{Department of Physics,
National University of Singapore,
3 Science Drive 2, Singapore 117543}

\author{Gerard John Wong Zhongwei}
\affiliation{Centre for Quantum Technologies,
National University of Singapore,
3 Science Drive 2, Singapore 117543}

\author{Rafael Rabelo}
\affiliation{Instituto de F\'isica Gleb Wataghin,
Universidade Estadual de Campinas (Unicamp),
Rua S\'ergio Buarque de Holanda 777,
Campinas, S\~ao Paulo 13083-859, Brazil}

\author{Gabriel Ruffolo}
\affiliation{Instituto de F\'isica Gleb Wataghin,
Universidade Estadual de Campinas (Unicamp),
Rua S\'ergio Buarque de Holanda 777,
Campinas, S\~ao Paulo 13083-859, Brazil}

\author{Pawe{\l} Kurzy\'nski}
\affiliation{Institute of Spintronics and Quantum Information,
Faculty of Physics, Adam Mickiewicz University,
Uniwersytetu Pozna\'nskiego 2, 61-614 Pozna\'n, Poland}

\author{Dagomir Kaszlikowski}
\affiliation{Centre for Quantum Technologies,
National University of Singapore,
3 Science Drive 2, Singapore 117543}
\affiliation{Department of Physics,
National University of Singapore,
3 Science Drive 2, Singapore 117543}

\date{}

\begin{abstract}
Can a joint measurement reveal nonclassical behaviour in independently
prepared systems that share no entanglement? We study network contextuality
by extending the two-qubit test of bi-contextuality [G.~Ruffolo et al.,
arXiv:2507.08461 (2025)] to
projective measurements with more than two outcomes. In one laboratory,
an observer measures the systems individually or jointly performs two
compatible global measurements constructed from them. The individual
probabilities constrain the possible classical correlations at each source.
For our preparations, every noncontextual model with outcome-deterministic
and source-independent assignments predicts a nonzero average product of
the two global outcomes.
The joint quantum measurement gives zero. We construct explicit product
states realizing this contradiction for $N=d$ independent sources, each
preparing one $d$-dimensional system, with $2\leq d\leq9$.
We derive an inequality in observed
probabilities, prove its single-source bounds are exact for these
preparations, and give sufficient thresholds for violations under independent
depolarizing noise. For this family of preparations the dimension range is
sharp: for every $d\geq10$ the individual probabilities admit a zero assigned
average, so the certificate fails. The same data admit a noncontextual explanation when source
independence is relaxed.
\end{abstract}

\maketitle

\section{Introduction}
\label{sec:introduction}

Can the measurement statistics of independently prepared systems be
explained by values carried by each system before measurement?
Bi-contextuality shows that the answer can be no even for two
unentangled qubits~\cite{Ruffolo2025}. Two sources send their systems
to one laboratory, where the experimenter measures them individually
or jointly performs compatible product measurements. No model in which
each system carries definite values and the sources are independent
reproduces the data. Allowing source correlations restores a model. We
call this failure of an independent-source noncontextual explanation
network contextuality.

The argument is short. Let $x_i,y_i$ be the values that system $i$
carries for two incompatible measurements, and let $c_i$ be the average
of $x_iy_i$ at source $i$. If the sources are independent, the average
product of the two global outcomes is
\begin{equation}
 C=\prod_i c_i .
\end{equation}
We choose preparations whose individual probabilities force every $c_i$
away from zero. Then $C\ne0$ in every such model, whereas the joint
quantum measurement gives $C=0$.

The original work treated two qubits. It gave a necessary and sufficient
criterion for the binary scenario and a proof-of-principle experimental
violation~\cite{Ruffolo2025}. Here we extend the construction to
projective measurements with $d$ outcomes. Our network has $N$
independent sources, each preparing one $d$-dimensional system. Thus $N$
counts the sources, while $d$ is the dimension of each system. The
measurements are unitary Weyl--Heisenberg operators whose eigenvalues
label the outcomes. Beyond binary outcomes these labels, and the
averages $c_i$, are complex. The central step is to exclude zero
throughout the allowed region of each $c_i$.

We realize the contradiction for $2\leq d\leq9$ using $N=d$ independent
sources and product states. We give explicit states, the exact lower
bound on each $|c_i|$, an inequality in observed probabilities, and
sufficient thresholds for independent depolarizing noise. For these
preparations the dimension range is sharp: for every $d\geq10$
the same state formula admits a zero assigned average at each source
(Appendix~\ref{app:general-weights}).

Section~\ref{sec:qubits} presents the whole argument for two qubits.
Sections~\ref{sec:experiment} and \ref{sec:classical} state the
experiment and the model for $N$ sources.
Sections~\ref{sec:construction}--\ref{sec:examples} give the construction
for $d$ outcomes, its proof, and explicit states.
Section~\ref{sec:noise} treats observed probabilities and noise.
The assumptions and limitations are collected in
Sec.~\ref{sec:discussion}.

\section{The argument for two qubits}
\label{sec:qubits}

Two independent sources each prepare a qubit in the state
\begin{equation}
 \ket{\psi_2}=\frac{\ket{0}+e^{-\ii\pi/4}\ket{1}}{\sqrt2}.
 \label{eq:qubit-state}
\end{equation}
On each qubit the experimenter can measure $X=\sigma_x$ or
$Y=\sigma_y$. With $Z=\sigma_z$, the state gives
\begin{equation}
 \langle X\rangle=\frac1{\sqrt2},\qquad
 \langle Y\rangle=-\frac1{\sqrt2},\qquad
 \langle Z\rangle=0.
 \label{eq:qubit-averages}
\end{equation}
These averages determine the binary outcome probabilities. The
experimenter can also jointly measure the products $A=X\otimes X$ and
$B=Y\otimes Y$. They commute, since $AB=BA=-Z\otimes Z$.

\emph{Classical account.} Suppose qubit $i$ carries values
$x_i,y_i=\pm1$, which the measurements reveal. Their joint distribution
$p_i(x_i,y_i)$ is unknown, but its marginals are the observed
probabilities. The products take the values $a=x_1x_2$ and $b=y_1y_2$,
whether they are measured jointly or through the individual outcomes.
If the sources are independent, the average of $ab$ factorizes:
\begin{equation}
 C=c_1c_2,\qquad c_i=\sum_{x_i,y_i}x_iy_i\,p_i(x_i,y_i).
\end{equation}

The individual statistics restrict each $c_i$. For every pair of
possible outcomes, positivity gives $(1-x_i)(1+y_i)\geq0$. Averaging
over a distribution $p_i$ with the required marginals yields, using the
measured averages in Eq.~\eqref{eq:qubit-averages},
\[
 1+\langle Y\rangle-\langle X\rangle-c_i\geq0,
\]
and hence
\begin{equation}
 c_i\leq1-\sqrt2<0.
 \label{eq:qubit-ci}
\end{equation}
Every allowed assignment therefore has $c_i<0$, with
$|c_i|\geq\sqrt2-1$. This bound can be attained: for example,
$p_i(+,+)=p_i(-,-)=(1-1/\sqrt2)/2$,
$p_i(+,-)=1/\sqrt2$, and $p_i(-,+)=0$ reproduce the measured
marginals and give $c_i=1-\sqrt2$. Independence then requires
\begin{equation}
 C=c_1c_2\geq(\sqrt2-1)^2>0.
\end{equation}

\emph{Quantum prediction.} For the product preparation, the joint
measurement gives
\begin{equation}
 C=\langle AB\rangle=-\langle Z\rangle^2=0.
\end{equation}

The two accounts disagree. This is the bi-contextuality of
Ref.~\cite{Ruffolo2025}, expressed with the present measurement and
state conventions. No entanglement is used. The contradiction
disappears when the sources may be correlated
(Appendix~\ref{app:joint-distribution}).

The rest of the paper repeats these three steps for $d$ outcomes: the
product formula $C=\prod_ic_i$, a bound keeping every $c_i$ away from
zero, and a state for which the quantum value of $C$ is zero.

\emph{More qubits.} For $N$ sources preparing the same state, define
\begin{equation}
 A=X^{\otimes N},\qquad B=Y^{\otimes N}.
\end{equation}
Since $XY=-YX$, we have $AB=(-1)^N BA$, so the joint setting is
available for even $N$. Each $c_i\leq1-\sqrt2<0$; independence therefore
requires
\begin{equation}
 C=\prod_{i=1}^N c_i\geq(\sqrt2-1)^N>0.
 \label{eq:qubit-bound}
\end{equation}
Quantum theory instead predicts
\begin{equation}
 C=\bigl(\bra{\psi_2}XY\ket{\psi_2}\bigr)^N
  =\bigl(\ii\bra{\psi_2}Z\ket{\psi_2}\bigr)^N=0.
 \label{eq:qubit-zero}
\end{equation}
Here $\langle XY\rangle$ calculates the quantum prediction and differs
from the assigned average $c_i$. For $N=2m$, the products decompose into
$m$ commuting two-qubit blocks, each repeating bi-contextuality.

\section{Experiment}
\label{sec:experiment}

We now state the experiment for $N$ sources and $d$ outcomes.
On each trial, $N$ independent sources each send one $d$-dimensional
system to a single laboratory. Each individual measurement box
$X^{(i)},Y^{(i)}$ has one system input, one system output, and one
classical outcome output. Each global box $A,B$ has $N$ system inputs
and outputs and one recorded outcome, $a$ or $b$, with operators
\begin{equation}
 A=X^{(1)}\otimes\cdots\otimes X^{(N)},\qquad
 B=Y^{(1)}\otimes\cdots\otimes Y^{(N)}.
 \label{eq:products}
\end{equation}
The experimenter independently chooses one of three settings:
\begin{enumerate}
\item Pass the systems through box $A$, record $a$, and then measure
$X^{(i)}$ on each output system, recording $x_1,\ldots,x_N$.
Check that $a=\prod_i x_i$ on every ideal trial.
\item Pass the systems through box $B$, record $b$, and then measure
$Y^{(i)}$ on each output system, recording $y_1,\ldots,y_N$.
Check that $b=\prod_i y_i$ on every ideal trial.
\item Pass the systems through the same box $A$ and then the same box
$B$, recording $(a,b)$ without resolving the individual outcomes.
\end{enumerate}
The preparation procedures are the same for all settings, and the choice
is independent of the sources. The two global devices are reused across
settings: $A$ in settings 1 and 3, and $B$ in settings 2 and 3.

Each individual measurement has $d$ rank-one orthogonal projectors summing
to the identity. The unitary operators $X^{(i)},Y^{(i)}$ specify these
projectors through their eigenbases; their distinct eigenvalues label
the outcomes. These labels may be complex, while probabilities remain real
and nonnegative.

The individual measurements are incompatible, but the product operators
are chosen so that $[A,B]=0$. We take each global box to be the
L\"uders instrument of the corresponding projective measurement. Let $\Pi_a^A$ denote the spectral projector of $A$
for outcome $a$, and similarly for $B$. On input $\rho$, the outcome-$a$
branch of box $A$ produces the unnormalized state
$\Pi_a^A\rho\Pi_a^A$, and analogously for $B$. A global box then leaves
the compatible individual statistics unchanged, and the $A$--$B$
sequence is a joint measurement of $A$ and $B$
(Appendix~\ref{app:measurements}).

\begin{figure}[htb]
\centering
\resizebox{0.95\linewidth}{!}{%
\begin{tikzpicture}[x=1cm,y=-1cm]
\begin{scope}
 \draw[panel] (0,0) rectangle (9.25,7.75);
 \node[title] at (.3,.43) {(a) Independent preparations};
 \node[subtitle] at (.3,.94) {$N$ sources, one system per source};
 \sources\laboratory
 \node[product,minimum width=72mm,minimum height=12mm] (choice) at (4.6,4.15)
  {Choose setting 1, 2, or 3\\[3pt]independently of the sources};
 \draw[qflow] (s1.south) -- (1.35,3.55);
 \draw[qflow] (s2.south) -- (3.5,3.55);
 \draw[qflow] (sn.south) -- (7.8,3.55);
 \draw[qflow] (1.35,4.75) -- (1.35,6.6);
 \draw[qflow] (3.5,4.75) -- (3.5,6.6);
 \draw[qflow] (7.8,4.75) -- (7.8,6.6);
 \node[note,align=center] at (4.6,7.35)
  {Repeat the same preparation procedures for all three settings.};
\end{scope}
\begin{scope}[shift={(9.65,0)}]
 \sequentialpanel{(b) Setting 1: $A$ then individual $X$}{A}{X}{a}
 \draw[cflow] (m1.east) -- +(0.65,0) node[above,font=\small] {$x_1$};
 \draw[cflow] (m2.east) -- +(0.65,0) node[above,font=\small] {$x_2$};
 \draw[cflow] (mn.east) -- +(0.45,0) node[above,font=\small] {$x_N$};
 \node[note,align=center] at (4.6,7.35)
  {Record $(a,x_1,\ldots,x_N)$;\;check $a=\prod_i x_i$.};
\end{scope}
\begin{scope}[shift={(0,8.15)}]
 \sequentialpanel{(c) Setting 2: $B$ then individual $Y$}{B}{Y}{b}
 \draw[cflow] (m1.east) -- +(0.65,0) node[above,font=\small] {$y_1$};
 \draw[cflow] (m2.east) -- +(0.65,0) node[above,font=\small] {$y_2$};
 \draw[cflow] (mn.east) -- +(0.45,0) node[above,font=\small] {$y_N$};
 \node[note,align=center] at (4.6,7.35)
  {Record $(b,y_1,\ldots,y_N)$;\;check $b=\prod_i y_i$.};
\end{scope}
\begin{scope}[shift={(9.65,8.15)}]
 \draw[panel] (0,0) rectangle (9.25,7.75);
 \node[title] at (.3,.43) {(d) Setting 3: $A$ then $B$};
 \node[subtitle] at (.3,.94) {Reuse the global boxes from settings 1 and 2};
 \sources\laboratory
 \node[measure,minimum width=72mm] (ga) at (4.6,3.1) {$A$};
 \node[measure,minimum width=72mm] (gb) at (4.6,5.1) {$B$};
 \draw[qflow] (s1.south) -- (1.35,2.7);
 \draw[qflow] (s2.south) -- (3.5,2.7);
 \draw[qflow] (sn.south) -- (7.8,2.7);
 \draw[qflow] (1.35,3.5) -- (1.35,4.7);
 \draw[qflow] (3.5,3.5) -- (3.5,4.7);
 \draw[qflow] (7.8,3.5) -- (7.8,4.7);
 \draw[qflow] (1.35,5.5) -- (1.35,6.6);
 \draw[qflow] (3.5,5.5) -- (3.5,6.6);
 \draw[qflow] (7.8,5.5) -- (7.8,6.6);
 \draw[cflow] (ga.east) -- (8.85,3.1) node[midway,above,font=\small] {$a$};
 \draw[cflow] (gb.east) -- (8.85,5.1) node[midway,above,font=\small] {$b$};
 \node[note,align=center] at (4.6,7.35)
  {Record $(a,b)$ to obtain $P_{AB}$; individual $x_i,y_i$ are not resolved.};
\end{scope}
\draw[qflow] (3.1,16.5) -- (3.9,16.5);
\node[note,anchor=west] at (4.05,16.5) {system input / output};
\draw[cflow] (10.1,16.5) -- (10.9,16.5);
\node[note,anchor=west] at (11.05,16.5) {recorded outcome};
\end{tikzpicture}}
\caption{Sequential measurement procedures for $N$ independent sources.
The schematic generalizes the two-source scenario of Ref.~\cite{Ruffolo2025}
to $N$ independent sources.
(a) The preparation is the same for all settings. (b) Box $A$ precedes
the individual $X^{(i)}$ measurements. (c) Box $B$ precedes the individual
$Y^{(i)}$ measurements. (d) The same $A,B$ boxes are used in sequence,
without resolving individual outcomes. Solid arrows carry systems;
dashed arrows carry recorded outcomes; ellipses omit system wires.
Each box retains its system outputs. The instruments and probability
checks are specified in the text.}
\label{fig:network-contextuality}
\end{figure}
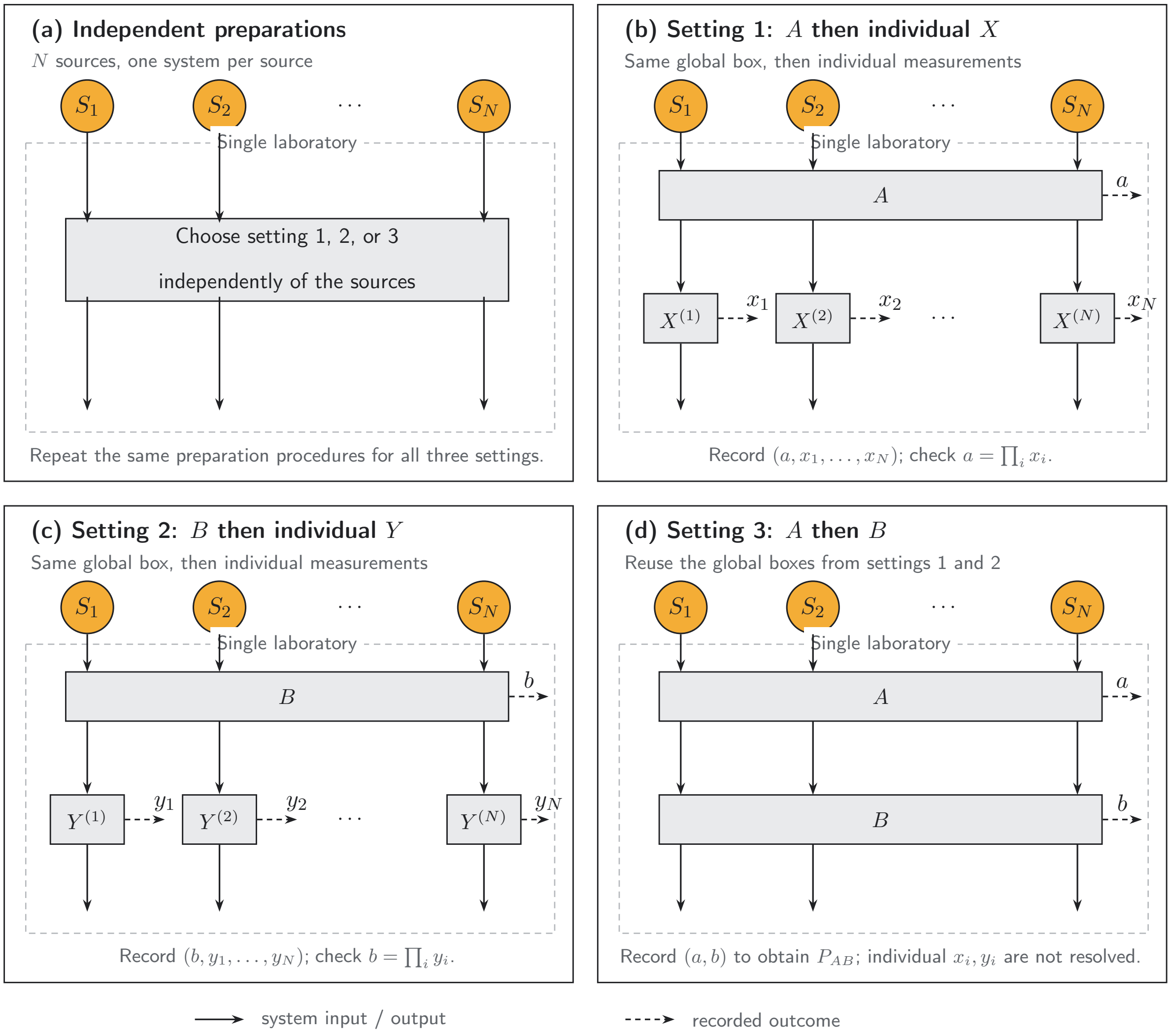

Figure~\ref{fig:network-contextuality} shows the procedures. Reusing the boxes
does not replace the probability checks below. In particular, $B$ receives
the outputs of $A$ in setting 3 but the initial inputs in setting 2, so
agreement of its marginals is a nontrivial no-disturbance check.

The raw data consist of
\begin{equation}
 P_{AX}(a,x_1,\ldots,x_N),\qquad
 P_{BY}(b,y_1,\ldots,y_N),\qquad P_{AB}(a,b).
 \label{eq:data}
\end{equation}
Retaining only the individual outcome lists defines
$P_X(\boldsymbol x)=\sum_aP_{AX}(a,\boldsymbol x)$ and
$P_Y(\boldsymbol y)=\sum_bP_{BY}(b,\boldsymbol y)$, where
$\boldsymbol x=(x_1,\ldots,x_N)$ and similarly for $\boldsymbol y$.
The ideal model requires trial-by-trial product agreement:
\begin{equation}
 P_{AX}(a,\boldsymbol x)=\delta_{a,\prod_i x_i}P_X(\boldsymbol x),\qquad
 P_{BY}(b,\boldsymbol y)=\delta_{b,\prod_i y_i}P_Y(\boldsymbol y).
 \label{eq:recorded-product-agreement}
\end{equation}
These relations are checkable because both the global outcome and the
individual list are recorded in settings 1 and 2.

The global marginal probabilities must also agree across settings:
\begin{align}
 \sum_{\substack{x_1,\ldots,x_N\\\prod_i x_i=a}}P_X(x_1,\ldots,x_N)
 &=\sum_bP_{AB}(a,b),\nonumber\\
 \sum_{\substack{y_1,\ldots,y_N\\\prod_i y_i=b}}P_Y(y_1,\ldots,y_N)
 &=\sum_aP_{AB}(a,b).
 \label{eq:agreement}
\end{align}
These no-disturbance conditions~\cite{Budroni2022} sum over lists with
the indicated product. Appendix~\ref{app:measurements} derives them and
the recorded product relations for every input state under the specified
L\"uders instruments. It also proves that $P_X,P_Y$ equal the direct
individual-measurement probabilities on the initial state. Source
independence concerns that initial preparation; conditioning on a global
outcome can produce an entangled state.

\section{The independent-source noncontextual model}
\label{sec:classical}

The model combines two assumptions. Noncontextuality requires the value
assigned to a product measurement to be the same in each compatible
context~\cite{KochenSpecker1967,Budroni2022}. Source independence
requires the assignments supplied by separate preparations to have a
factorized probability distribution, as in quantum
networks~\cite{Branciard2010,Branciard2012,Tavakoli2022}. Our network
consists of independent sources feeding a single measurement laboratory;
spatial separation, as in Bell experiments, is not required~\cite{Brunner2014}.

We use the deterministic-assignment notion of
noncontextuality~\cite{AbramskyBrandenburger2011}. A model is specified
by a normalized, nonnegative joint distribution
$p(x_1,y_1,\ldots,x_N,y_N)$, whose realizations assign a definite pair
$(x_i,y_i)$ to each system. Each pair is assumed to depend only on its
own source's hidden variable. The probabilities describe randomness in
these assignments across trials; outcome determinism and this
source-independent assignment rule are assumptions of the model.
Noncontextuality
requires the value assigned to $A$, or to $B$, to be independent of
which of its two measurement procedures is used. The joint setting
must therefore return the same products $a=\prod_i x_i$ and
$b=\prod_i y_i$ as the individual settings.

Independence of the source variables imposes the additional
requirement~\cite{Branciard2010}
\begin{equation}
 p(x_1,y_1,\ldots,x_N,y_N)=\prod_{i=1}^N p_i(x_i,y_i),
 \label{eq:independent-pairs}
\end{equation}
where each $p_i$ is normalized and nonnegative. The setting choice is
assumed to be independent of the source variables. Correlations between $x_i$ and
$y_i$ within one source remain unrestricted. Identical quantum
preparations do not require identical distributions $p_i$.

The model must reproduce all three observed distributions:
\begin{equation}
 \begin{aligned}
 P_X(x_1,\ldots,x_N)&=\prod_{i=1}^N\sum_{y_i}p_i(x_i,y_i),\\
 P_Y(y_1,\ldots,y_N)&=\prod_{i=1}^N\sum_{x_i}p_i(x_i,y_i),
 \end{aligned}
 \label{eq:individual-model-distributions}
\end{equation}
The raw individual-setting distributions are then given by
Eq.~\eqref{eq:recorded-product-agreement}. For the third setting,
\begin{equation}
 P_{AB}(a,b)=
 \sum_{\substack{x_1,y_1,\ldots,x_N,y_N\\
                 \prod_i x_i=a,\;\prod_i y_i=b}}
 \prod_{i=1}^N p_i(x_i,y_i).
 \label{eq:joint-model-distribution}
\end{equation}
All sums run over measurement outcomes. Data excluding every such model
display network contextuality with $N$ independent sources. The two-qubit
case with $N=2$ is bi-contextuality.

Consider the measured product average
\begin{equation}
 C=\langle AB\rangle=\sum_{a,b}ab\,P_{AB}(a,b)
 \label{eq:measured-product-average}
\end{equation}
and the assigned average for each source,
\begin{equation}
 c_i=\sum_{x_i,y_i}x_i y_i\,p_i(x_i,y_i).
 \label{eq:assigned-pair-average}
\end{equation}
The experiment measures $C$. The average $c_i$ concerns values assigned
to incompatible measurements, and is constrained by their observed
probabilities through the marginals of $p_i$. Using the same product
values across settings and applying source independence gives
\begin{equation}
 C=\sum_{x_1,y_1,\ldots,x_N,y_N}
       \prod_{i=1}^N[x_i y_i p_i(x_i,y_i)]
   =\prod_{i=1}^N c_i.
 \label{eq:factorized-product-average}
\end{equation}
Thus a classical prediction $C=0$ requires $c_i=0$ at some source.
Our preparations exclude this possibility at every source, although
the joint quantum measurement gives zero. Appendix~\ref{app:general-weights}
characterizes the allowed assigned averages.

For this three-setting experiment, the no-disturbance conditions
\eqref{eq:agreement}, together with the recorded product relations
\eqref{eq:recorded-product-agreement}, suffice to construct a classical
joint distribution reproducing all three raw setting distributions
without source independence. Appendix~\ref{app:joint-distribution} gives
it explicitly, generalizing the two-source construction of
Ref.~\cite{Ruffolo2025} and using the compatible-marginal gluing method
discussed in Ref.~\cite{Ramanathan2012}.

\section{Beyond binary outcomes}
\label{sec:construction}

From here on the number of sources equals the dimension of each system,
$N=d$. This is the smallest number for which the two global measurements
commute, as shown below. The classical bound of Sec.~\ref{sec:bound}
holds for $2\leq d\leq9$.

\subsection{Measurements}

On each $d$-dimensional system, the experimenter chooses between two
measurements, $X$ and $Y$, defined below. Each is a projective
measurement in an orthonormal basis, with one outcome per basis vector.

In the basis $\ket{0},\ldots,\ket{d-1}$, set
$\omega=e^{2\pi\ii/d}$ and define
\begin{equation}
 X\ket{q}=\ket{q+1},\qquad Z\ket{q}=\omega^q\ket{q},\qquad Y=XZ,
 \label{eq:weyl-definition}
\end{equation}
with addition modulo $d$~\cite{Schwinger1960,Vourdas2004}.
The operator $X$ shifts each basis vector to the next one, $Z$
attaches a phase, and $Y$ applies both operations. They obey
\begin{equation}
 ZX=\omega XZ,\qquad XY=\omega^{-1}YX.
 \label{eq:weyl-relation}
\end{equation}
Multiplying $Y$ by a fixed phase changes its outcome labels but
preserves its projectors and this commutation relation. In the
qubit convention of Sec.~\ref{sec:qubits}, $Y=\ii XZ=\sigma_y$.
For $d>2$, these operators are not Hermitian: their eigenvalues
label the outcomes of projective measurements in their eigenbases.
Complex averages are obtained by averaging the real and imaginary
parts of these labels separately.

For the operators in Eq.~\eqref{eq:weyl-definition}, the identities
$X^d=I$ and $(XZ)^d=(-1)^{d-1}I$, together with the Weyl relation
in Eq.~\eqref{eq:weyl-relation}, imply that each spectrum consists of
$d$ distinct, equally spaced points on the unit circle. Choose reference eigenvalues
$u$ of $X$ and $v$ of $Y$ to identify the favored outcomes in their
individual distributions. These are outcome labels. For $Y=XZ$,
$u^d=1$ and $v^d=(-1)^{d-1}$; the phase-adjusted qubit convention
$Y=\sigma_y$ instead has $v=\pm1$.
The possible labels are
\begin{equation}
 x\in\{u\omega^j:j=0,\ldots,d-1\},\qquad
 y\in\{v\omega^k:k=0,\ldots,d-1\}.
 \label{eq:eigenvalues}
\end{equation}
Write $\ket{x;X}$ for the normalized eigenvector of $X$ with eigenvalue
$x$, and $\ket{y;Y}$ for the normalized eigenvector of $Y$ with eigenvalue
$y$. The letters after the semicolon identify the measurement basis.
An eigenstate of either measurement gives uniform outcome
probabilities for the other. Equivalently, the bases are mutually
unbiased:
\begin{equation}
 \left|\langle x;X\mid y;Y\rangle\right|^2=\frac1d.
 \label{eq:overlaps}
\end{equation}
Appendix~\ref{app:state} derives this relation from
Eq.~\eqref{eq:weyl-relation}.

\subsection{State preparation}

Each source prepares the same pure state. We choose it so that the
quantum average of $XY$ vanishes, while the individual measurements
have biased outcome probabilities. Choose the phases of the two
eigenvectors so that $\langle u;X\mid v;Y\rangle=1/\sqrt d$,
and set
\begin{equation}
 \ket{\psi_d}=
 \frac{\ket{u;X}+e^{\ii(\pi/2-\pi/d)}\ket{v;Y}}{\sqrt{2+2h_d}},
 \qquad h_d=\frac{\sin(\pi/d)}{\sqrt d}.
 \label{eq:state}
\end{equation}
The relative phase makes the two contributions to the product
average cancel, giving
\begin{equation}
 \bra{\psi_d}XY\ket{\psi_d}=0.
 \label{eq:XY-zero}
\end{equation}
Appendix~\ref{app:state} verifies the normalization and cancellation.

The ideal quantum probabilities are
$q_{X,d}(x)=|\langle x;X\mid\psi_d\rangle|^2$ and
$q_{Y,d}(y)=|\langle y;Y\mid\psi_d\rangle|^2$.
Using Eq.~\eqref{eq:overlaps} gives
\begin{equation}
 \begin{split}
 \beta_d&=\frac{1+2h_d}{2+2h_d},\\
 q_{X,d}(x)&=\frac{1-\beta_d}{d}+\beta_d\delta_{x,u},\\
 q_{Y,d}(y)&=\frac{1-\beta_d}{d}+\beta_d\delta_{y,v}.
 \end{split}
 \label{eq:probabilities}
\end{equation}
The selected outcome has probability $(1-\beta_d)/d+\beta_d$; each
other outcome has probability $(1-\beta_d)/d$.
For qubits, choosing $u=+1$, $v=-1$ with $Y=\sigma_y$ recovers
Eq.~\eqref{eq:qubit-state} up to an overall phase. This choice
also explains the negative sign of the qubit assigned average:
$c_i/(uv)=-c_i$.

\subsection{Joint measurement}

Use the same $X,Y$ on every system and define $A=X^{\otimes N}$,
$B=Y^{\otimes N}$. Then
\begin{equation}
 AB=\omega^{-N}BA,
 \label{eq:commutation}
\end{equation}
so $A$ and $B$ commute precisely when $N$ is a multiple of $d$.
The smallest choice is $N=d$. For the product preparation
\begin{equation}
 \rho=\bigl(\ket{\psi_d}\bra{\psi_d}\bigr)^{\otimes d},
 \label{eq:product-state}
\end{equation}
the joint measurement gives
\begin{equation}
 C=\Tr(\rho AB)=\bigl(\bra{\psi_d}XY\ket{\psi_d}\bigr)^d=0.
 \label{eq:quantum-zero}
\end{equation}

\subsection{The basic case $N=d$ and repetitions with $N=md$}
\label{sec:basic-source-number}

The basic result uses $N=d$ independent sources, each supplying one
$d$-dimensional system. For $2\leq d\leq9$, the next section proves that
their individual probabilities require
\begin{equation}
 |C|\geq\Delta_d^d>0
 \qquad\text{in every independent-source noncontextual model},
 \label{eq:basic-source-bound}
\end{equation}
where $\Delta_d$ is the exact single-source bound derived in
Eq.~\eqref{eq:delta}. The joint quantum measurement gives $C=0$.
The case $d=2$ recovers bi-contextuality.

For $N=md$, the same products commute and the same argument gives
$|C|\geq\Delta_d^{md}>0$ against the quantum value zero. This repeats the
basic test on $m$ blocks of $d$ systems. It does not establish genuine
$md$-partite contextuality, which would require a separate definition and
proof even at $N=d$. Appendix~\ref{app:blocks} describes the block
structure and shows that, for other preparations, discarding block labels
can hide a block violation.

\section{Why the classical average cannot vanish}
\label{sec:bound}

\subsection{The restriction at each source}

Each classical distribution $p_i(x_i,y_i)$ must reproduce the
individual quantum probabilities:
\begin{equation}
 \sum_{y_i}p_i(x_i,y_i)=q_{X,d}(x_i),\qquad
 \sum_{x_i}p_i(x_i,y_i)=q_{Y,d}(y_i).
\end{equation}
The two marginals leave the correlation between the assigned values
unrestricted. We must therefore rule out a zero product average for
every nonnegative joint distribution with these marginals.

As in the qubit example, we seek a function that is nonnegative for every
allowed outcome pair. For any $d\geq2$, define
\begin{equation}
 D_d(x,y)=1+\Repart\frac{xy}{uv}
             -\Repart\frac{x}{u}-\Repart\frac{y}{v}.
 \label{eq:D}
\end{equation}
For $d=2,3,4$, $D_d(x,y)\geq0$ for all pairs; the
finite check is given in Appendix~\ref{app:weights}. Its average
with any $p_i$ must therefore be nonnegative.

The individual averages in Eq.~\eqref{eq:probabilities} obey
\begin{equation}
 \sum_x\frac{x}{u}q_{X,d}(x)=\beta_d,\qquad
 \sum_y\frac{y}{v}q_{Y,d}(y)=\beta_d.
\end{equation}
Thus averaging Eq.~\eqref{eq:D} gives
\begin{equation}
 0\leq1+\Repart\frac{c_i}{uv}-2\beta_d,
 \qquad |c_i|\geq2\beta_d-1>0.
 \label{eq:simple-bound}
\end{equation}
Dividing by $uv$ only rotates $c_i$ in the complex plane, since
$|uv|=1$. The inequality forces the real part of the rotated
average to be positive. It therefore excludes $c_i=0$, even when
the assigned average is complex.

\subsection{More outcomes and the global contradiction}

For $5\leq d\leq9$, some values of $D_d$ are negative. We add
nonnegative terms depending on one outcome at a time:
\begin{equation}
 W_d(u\omega^j,v\omega^k)
 =D_d(u\omega^j,v\omega^k)+\ell_{d,j}+\ell_{d,k}\geq0.
 \label{eq:weights}
\end{equation}
The coefficients $\ell_{d,j}$ are fixed numbers for each dimension,
not properties of the unknown classical distribution. Their explicit
values and the proof of nonnegativity are given in
Appendix~\ref{app:weights}. The selected outcomes need no correction,
so $\ell_{d,0}=0$. For $d=2,3,4$, every coefficient is zero.

Because the added terms depend on single outcomes, their averages
are fixed by the measured individual probabilities. Define
\begin{equation}
 K_d=\frac2d\sum_{j=1}^{d-1}\ell_{d,j}.
 \label{eq:optimal-K}
\end{equation}
Each nonselected outcome has probability $(1-\beta_d)/d$.
Averaging $W_d$ therefore yields
\begin{equation}
 0\leq1+\Repart\frac{c_i}{uv}-2\beta_d+K_d(1-\beta_d).
 \label{eq:weighted-bound}
\end{equation}
It follows that every source obeys
\begin{equation}
 |c_i|\geq\Delta_d,
 \qquad
 \Delta_d=2\beta_d-1-K_d(1-\beta_d)
          =\frac{2h_d-K_d}{2+2h_d}.
 \label{eq:delta}
\end{equation}
The bound $\Delta_d$ is strictly positive for $2\leq d\leq9$.
Appendix~\ref{app:weights} proves this and constructs assigned
joint distributions attaining it. Thus $\Delta_d$ is the exact
smallest magnitude allowed by the ideal individual probabilities.
Appendix~\ref{app:general-weights} explains how to find this minimum
directly, without guessing the auxiliary function. It also proves that
the range cannot be extended for the states of Eq.~\eqref{eq:state}:
for every $d\geq10$ some allowed distribution has $c_i=0$.

Source independence now completes the argument:
\begin{equation}
 \boxed{|C|=\prod_{i=1}^d|c_i|\geq\Delta_d^d>0,
 \qquad N=d,\quad 2\leq d\leq9.}
 \label{eq:main-bound}
\end{equation}
The joint quantum prediction is $C=0$, violating this bound. For $N=md$
the same steps give $|C|\geq\Delta_d^{md}$. The inequality tests one average of $P_{AB}$; reproducing
the full distribution imposes further constraints on a classical model.

\section{Explicit states}
\label{sec:examples}

The qubit state is given in Eq.~\eqref{eq:qubit-state}.
For three and four outcomes, the following vectors give the
states of Eq.~\eqref{eq:state}, up to an overall phase.
Entries are ordered in the computational basis
$\ket{0},\ldots,\ket{d-1}$.

For qutrits, use $Y=XZ$, $\omega=e^{2\pi\ii/3}$,
and selected eigenvalues $u=\omega$, $v=\omega^2$.
Each source prepares
\begin{equation}
 \ket{\psi_3}
 =\frac13
 \begin{pmatrix}
 -\omega\\
 2\omega\\
 2
 \end{pmatrix}.
 \label{eq:qutrit}
\end{equation}
Here $\beta_3=2/3$, and every classical assigned-pair
distribution reproducing the individual probabilities satisfies
$|c_i|\geq1/3$. The joint quantum measurement gives $C=0$
for $N=3,6,9,\ldots$.

For ququarts, use $Y=XZ$ and $\omega=\ii$.
Since $Y^4=-I$, its eigenvalues are
$e^{\ii\pi/4}\ii^k$, for $k=0,1,2,3$.
Select $u=\ii$ and $v=e^{-\ii\pi/4}$.
Set $s=(\ii+e^{3\ii\pi/4})/2$ and prepare
\begin{equation}
 \ket{\psi_4}
 =\sqrt{\frac{4-\sqrt2}{7}}
 \begin{pmatrix}
 -s\\
 0\\
 s\\
 1
 \end{pmatrix}.
 \label{eq:ququart}
\end{equation}
Here $\beta_4=(3+\sqrt2)/7$ and
$|c_i|\geq(2\sqrt2-1)/7$. The contradiction holds for
$N=4,8,12,\ldots$.

For comparison, the three simplest bounds are
\begin{center}
\begin{tabular}{ccc}
\toprule
Dimension $d$ & Probability parameter $\beta_d$
& Single-source bound $\Delta_d$\\
\midrule
$2$ & $1/\sqrt2$ & $\sqrt2-1$\\
$3$ & $2/3$ & $1/3$\\
$4$ & $(3+\sqrt2)/7$ & $(2\sqrt2-1)/7$\\
\bottomrule
\end{tabular}
\end{center}
The individual probabilities follow from Eq.~\eqref{eq:probabilities}.

Appendix~\ref{app:state} gives every computational-basis component of
the states for $d=5,\ldots,9$. Their bounds and noise thresholds are
listed in Sec.~\ref{sec:noise}.

\section{An inequality in measured probabilities and noise}
\label{sec:noise}

\subsection{Using the measured individual probabilities}

An experiment need not produce the ideal probabilities exactly.
Let $q_{X,i}(x_i)$ and $q_{Y,i}(y_i)$ be the observed probabilities
at source $i$, obtained by summing the first two settings' outcome
lists over the other systems:
\begin{equation}
 \begin{aligned}
 q_{X,i}(x_i)&=\sum_{\{x_j:j\ne i\}}P_X(x_1,\ldots,x_N),\\
 q_{Y,i}(y_i)&=\sum_{\{y_j:j\ne i\}}P_Y(y_1,\ldots,y_N).
 \end{aligned}
\end{equation}
Any classical explanation must reproduce these same probabilities.
Averaging the nonnegative function $W_d$ with these marginals gives
$\Repart[c_i/(uv)]\geq b_i$, where
\begin{equation}
 \begin{split}
 b_i={}&\sum_{j=0}^{d-1}
 \left[\cos\frac{2\pi j}{d}-\ell_{d,j}\right]q_{X,i}(u\omega^j)\\
 &+\sum_{k=0}^{d-1}
 \left[\cos\frac{2\pi k}{d}-\ell_{d,k}\right]q_{Y,i}(v\omega^k)-1.
 \end{split}
 \label{eq:Bi}
\end{equation}
Here the coefficients are those proved valid for $2\leq d\leq9$
in Appendix~\ref{app:weights}. Since
$|c_i|\geq\max\{0,b_i\}$, independence requires
\begin{equation}
 \boxed{|C|\geq\prod_{i=1}^N\max\{0,b_i\}.}
 \label{eq:measured-bound}
\end{equation}
The first two settings determine $b_i$ and the third determines $C$.
A violation excludes the model of Sec.~\ref{sec:classical}; ideally,
$b_i=\Delta_d$. If any $b_i\leq0$, the certificate becomes trivial.
For example, a maximally mixed source gives $b_i=-(1+K_d)<0$, and
the noisy family below gives $b_i\leq0$ for $t\leq t_d^{\mathrm{crit}}$.
Other tests may still detect contextuality. Appendix~\ref{app:general-weights}
gives the exact bound from arbitrary individual marginals.

\subsection{Independent depolarizing noise}

For a specific noise model, suppose each source independently
prepares
\begin{equation}
 \rho_d(t)
 =t\ket{\psi_d}\bra{\psi_d}+(1-t)\frac{I}{d},
 \qquad 0\leq t\leq1.
 \label{eq:noise-state}
\end{equation}
The total state is $\rho_d(t)^{\otimes N}$. Independence also permits
nonidentical product inputs $\rho_1\otimes\cdots\otimes\rho_N$:
Eq.~\eqref{eq:measured-bound} applies with source-dependent probabilities,
and Appendix~\ref{app:general-weights} gives their exact bounds. Here we
analyze the symmetric family, whose individual probabilities are
\[
 q_{X,i}(x_i)
 =\frac{1-t\beta_d}{d}+t\beta_d\delta_{x_i,u},
 \qquad
 q_{Y,i}(y_i)
 =\frac{1-t\beta_d}{d}+t\beta_d\delta_{y_i,v}.
\]
Cyclicity of the trace and Eq.~\eqref{eq:weyl-relation} give
\[
 \Tr(XY)=\Tr(YX)=\omega\Tr(XY).
\]
Since $\omega\ne1$, $\Tr(XY)=0$.
Together with Eq.~\eqref{eq:XY-zero}, this gives
$\Tr[\rho_d(t)XY]=0$.
Thus the joint measurement still predicts $C=0$ when $N$
is a multiple of $d$.

Substitution of the noisy probabilities into Eq.~\eqref{eq:Bi}
gives, for $2\leq d\leq9$,
\begin{equation}
 b_i=(2+K_d)t\beta_d-(1+K_d).
 \label{eq:noisy-source-bound}
\end{equation}
Since $C=0$, the inequality is violated whenever the common $b_i>0$:
\begin{equation}
 t>t_d^{\mathrm{crit}}=\frac{1+K_d}{(2+K_d)\beta_d}.
 \label{eq:threshold}
\end{equation}
The sufficient thresholds for qubits, qutrits, and ququarts are
\[
 \frac1{\sqrt2},\qquad
 \frac34,\qquad
 \frac{3-\sqrt2}{2},
\]
respectively. For the chosen noisy states, these thresholds are exact
for excluding a zero single-source product average, as proved in
Appendix~\ref{app:weights}. They remain sufficient thresholds for the
full experiment: below them, a model for $P_{AB}$ need not exist.
The measurements are assumed ideal.

The size of the violation matters for an experiment. At the smallest
allowed number of sources, $N=d$, the difference between the lower
bound in Eq.~\eqref{eq:main-bound} and the quantum value $|C|=0$ is
\begin{center}
\begin{tabular}{cccc}
\toprule
$d=N$ & \shortstack{Single-source bound\\$\Delta_d$}
& \shortstack{Global bound\\$\Delta_d^d$} & $t_d^{\mathrm{crit}}$\\
\midrule
$2$ & $0.414214$ & $1.72\times10^{-1}$ & $0.707107$\\
$3$ & $0.333333$ & $3.70\times10^{-2}$ & $0.750000$\\
$4$ & $0.261204$ & $4.65\times10^{-3}$ & $0.792893$\\
$5$ & $0.140518$ & $5.48\times10^{-5}$ & $0.892844$\\
$6$ & $0.100314$ & $1.02\times10^{-6}$ & $0.920824$\\
$7$ & $0.066774$ & $5.92\times10^{-9}$ & $0.946120$\\
$8$ & $0.027962$ & $3.74\times10^{-13}$ & $0.977360$\\
$9$ & $0.014629$ & $3.07\times10^{-17}$ & $0.987910$\\
\bottomrule
\end{tabular}
\end{center}
The gap shrinks rapidly with dimension and further as $\Delta_d^N$
with source number. Certification requires the upper allowed value of
$|C|$, accounting for statistical and systematic errors in its real and
imaginary parts, to lie below a lower confidence bound on the right-hand
side of Eq.~\eqref{eq:measured-bound}. The tiny higher-dimensional gaps
therefore demand very small absolute errors. The noise thresholds assume
ideal measurements; a finite-statistics and measurement-error analysis
remains open.

\section{Discussion}
\label{sec:discussion}

We extend bi-contextuality to projective measurements with more than two
outcomes. For $N=d$ independent sources of $d$-dimensional systems,
$2\leq d\leq9$, the individual probabilities exclude a zero assigned
average at every source. Independence then conflicts with the zero global
quantum average. The argument accommodates complex outcome labels and
uses product preparations throughout.

\emph{Assumptions.} A violation of Eq.~\eqref{eq:measured-bound} rules
out the independent-source noncontextual model of
Sec.~\ref{sec:classical}. The interpretation assumes independent sources,
a setting choice independent of those sources, and the same
product measurements in the individual and joint settings. For
imperfect devices, agreement of the observed product probabilities is
necessary but does not by itself establish this last requirement.
Allowing source correlations restores a noncontextual model
(Appendix~\ref{app:joint-distribution}), locating the conflict in the
independent-source constraint. The test applies to a joint-measurement
node supplied by independent sources~\cite{Ruffolo2025,Tavakoli2022},
without entanglement between the incoming systems.

\emph{Limitations.} The cases $N=md$ repeat the basic construction and
do not establish genuine multipartite contextuality
(Sec.~\ref{sec:basic-source-number}). The inequality tests one average of
$P_{AB}$. The small higher-dimensional gaps limit experimental usefulness.
Because the single-source bounds are attained, further optimization of
the same average with the same marginals cannot enlarge them. Larger
violations require different preparations, measurements, or tests using
more of $P_{AB}$. Nonidentical product inputs are also admissible; their
potential advantage is not established by the symmetric examples here.

\emph{Dimension range.} The state formula extends beyond $d=9$, but for
every $d\geq10$ its individual probabilities admit a zero-average
coupling (Appendix~\ref{app:cutoff}). The range $2\leq d\leq9$ is
therefore sharp for this preparation and this certificate. The result
does not supply a model for the full $P_{AB}$ or establish a universal
dimension limit for network contextuality. A necessary and sufficient
higher-dimensional criterion remains open.

\begin{acknowledgments}
OpenAI Codex was used to assist with language editing, LaTeX
preparation, and numerical consistency checks.

GR acknowledges support from the Brazilian National Council
for Scientific and Technological Development (CNPq)
(Grant Number 150695/2026-8). RR acknowledges support from the Brazilian National Council
for Scientific and Technological Development (CNPq)
(Grant Number 316657/2023-9). DK acknowledges support from the National Research Foundation,
Singapore through the National Quantum Office, hosted in
A*STAR, under its Centre for Quantum Technologies Funding
Initiative (S24Q2d0009) and Quantum Engineering Programme
(W21Qpd0208).
\end{acknowledgments}

\section*{Data availability}
The analytical results supporting this work are contained in the
article and its appendices. No experimental data are reported.

\appendix

\section{Derivation of the prepared state and its probabilities}
\label{app:state}

From $YX=\omega XY$, applying $X$ to a $Y$ eigenvector
with eigenvalue $y$ produces a $Y$ eigenvector with eigenvalue
$\omega y$. Successive applications generate all $d$ distinct
eigenvalues, and the normalized vectors form an orthonormal
eigenbasis.

For a fixed $X$ eigenvector,
\begin{equation}
 \bra{x;X}X=x\,\bra{x;X}.
\end{equation}
Applying $X$ to the $Y$ eigenvectors therefore changes their
overlaps with this fixed vector only by a phase.
All $d$ squared overlap magnitudes are equal and sum to one,
proving Eq.~\eqref{eq:overlaps}.

Choose the phases of $\ket{u;X},\ket{v;Y}$ so that their
overlap is $1/\sqrt d$. For an arbitrary relative phase $\phi$,
define
\begin{equation}
 \ket{\psi(\phi)}
 =
 \frac{\ket{u;X}+e^{\ii\phi}\ket{v;Y}}
 {\sqrt{2+2\cos\phi/\sqrt d}}.
\end{equation}
The vector $Y\ket{u;X}$ has $X$ eigenvalue $\omega^{-1}u$,
different from $u$, and $X\ket{v;Y}$ has $Y$ eigenvalue
$\omega v$, different from $v$. It follows that
$\bra{u;X}XY\ket{u;X}=0$ and
$\bra{v;Y}XY\ket{v;Y}=0$. The cross terms are
\begin{align}
 \bra{u;X}XY\ket{v;Y}
 &=uv\,\langle u;X\mid v;Y\rangle
 =\frac{uv}{\sqrt d},\\
 \bra{v;Y}XY\ket{u;X}
 &=\omega^{-1}uv\,\langle v;Y\mid u;X\rangle
 =\frac{\omega^{-1}uv}{\sqrt d}.
\end{align}
Combining the two cross terms gives
\begin{equation}
 \bra{\psi(\phi)}XY\ket{\psi(\phi)}
 =\frac{uv(e^{\ii\phi}+\omega^{-1}e^{-\ii\phi})}
 {\sqrt d(2+2\cos\phi/\sqrt d)}.
\end{equation}
The numerator is $2uv e^{-\ii\pi/d}\cos(\phi+\pi/d)$ and vanishes
at $\phi_d=\pi/2-\pi/d$.
At this phase the squared norm of the unnormalized superposition is
$2+2h_d$.

For the $X$ measurement, the amplitudes are
\begin{align}
 \langle u;X\mid\psi_d\rangle
 &=\frac{1+e^{\ii\phi_d}/\sqrt d}{\sqrt{2+2h_d}},\\
 \langle x;X\mid\psi_d\rangle
 &=\frac{
 e^{\ii\phi_d}\,\langle x;X\mid v;Y\rangle}
 {\sqrt{2+2h_d}},
 \qquad x\ne u.
\end{align}
For the $Y$ measurement they are
\begin{align}
 \langle v;Y\mid\psi_d\rangle
 &=\frac{1/\sqrt d+e^{\ii\phi_d}}{\sqrt{2+2h_d}},\\
 \langle y;Y\mid\psi_d\rangle
 &=\frac{\langle y;Y\mid u;X\rangle}
 {\sqrt{2+2h_d}},
 \qquad y\ne v.
\end{align}
Taking squared magnitudes gives Eq.~\eqref{eq:probabilities}.
Since the eigenvalues in each spectrum sum to zero, these
probabilities also give
$\langle X\rangle=u\beta_d$ and
$\langle Y\rangle=v\beta_d$.

\subsection*{Components for $d=5,\ldots,9$}

For $d=5,\ldots,9$, one can specify every computational-basis component
without diagonalizing the operators. Set
\begin{equation}
 u=1,\qquad v=e^{\ii\pi(d-1)/d},\qquad
 g_{d,q}=e^{\ii\pi q(q-d)/d},\quad q=0,\ldots,d-1.
\end{equation}
An $X$ eigenvector at $u$ has all components $1/\sqrt d$. A $Y$ eigenvector
at $v$ has components $\kappa_d g_{d,q}/\sqrt d$, where the phase
$\kappa_d$ makes its overlap with the selected $X$ vector positive.
The phases and the resulting exact component formula are
\begin{center}
\begin{tabular}{cccccc}
\toprule
 $d$&$5$&$6$&$7$&$8$&$9$\\
\midrule
 $\kappa_d$&$-1$&$e^{5\ii\pi/4}$&$-\ii$&$e^{7\ii\pi/4}$&$1$\\
\bottomrule
\end{tabular}
\end{center}
\begin{equation}
 \langle q\mid\psi_d\rangle
 =\frac{1+e^{\ii(\pi/2-\pi/d)}\kappa_d g_{d,q}}
 {\sqrt{d(2+2h_d)}}.
 \label{eq:remaining-states}
\end{equation}
For these five dimensions the finite sums satisfy
$\kappa_d\sum_q g_{d,q}=\sqrt d$, fixing the required overlap phase.
The recurrence $g_{d,q}/g_{d,q-1}=v^{-1}\omega^{q-1}$ verifies the
$Y$ eigenvalue directly. Substitution into Eq.~\eqref{eq:state} gives
Eq.~\eqref{eq:remaining-states}.

\section{Proof and optimality of the probability bound}
\label{app:weights}

\subsection{Nonnegative weights}

This appendix proves the nonnegative probability bounds used in
Sec.~\ref{sec:bound} and shows that they can be attained. Write
$D^{(d)}_{jk}=D_d(u\omega^j,v\omega^k)$, with $D_d$ defined in
Eq.~\eqref{eq:D}. For $d=2,\ldots,9$, use coefficients
$\ell_{d,0}=0$ and $\ell_{d,d-j}=\ell_{d,j}$ for
$1\leq j\leq d-1$. For $d=2,3,4$, all coefficients vanish.
For the remaining dimensions the only potentially nonzero entries
are $j=1,2$ and their reflections:
\begin{equation}
\begin{array}{c|cc}
 d&\ell_{d,1}&\ell_{d,2}\\ \hline
 5,6&-D^{(d)}_{11}/2&0\\
 7,8&-D^{(d)}_{11}/2&D^{(d)}_{11}/2-D^{(d)}_{12}\\
 9&-D^{(d)}_{13}&D^{(d)}_{13}-D^{(d)}_{12}
\end{array}
 \label{eq:optimal-corrections}
\end{equation}
All other entries are zero. Direct summation gives
$2\sum_j\ell_{d,j}/d=K_d$ with $K_d$ from
Eq.~\eqref{eq:optimal-K}. We now verify both nonnegativity of the
weights and attainability of the resulting average.

For $\theta=2\pi j/d$ and $\eta=2\pi k/d$,
\begin{equation}
 D^{(d)}_{jk}=-4\cos\frac{\theta+\eta}{2}
              \sin\frac\theta2\sin\frac\eta2.
\end{equation}
The entries with $j=0$ or $k=0$ vanish. For nonzero indices a negative
entry requires $j+k<d/2$ or $j+k>3d/2$. Complex conjugation maps the
second set to the first. Thus it suffices to check the following
pairs and their transposes:
\begin{equation}
\begin{array}{c|l}
 d& (j,k)\text{ with }j+k<d/2\\ \hline
 2,3,4&\text{none}\\
 5,6&(1,1)\\
 7,8&(1,1),(1,2)\\
 9&(1,1),(1,2),(1,3),(2,2).
\end{array}
\end{equation}
For $d=5,6$, $D^{(d)}_{11}+2\ell_{d,1}=0$. For $d=7,8$, both
$D^{(d)}_{11}+2\ell_{d,1}=0$ and $D^{(d)}_{12}+\ell_{d,1}+\ell_{d,2}=0$.
The coefficients in Eq.~\eqref{eq:optimal-corrections} are nonnegative.
For $d=9$, the $(1,2)$ and $(1,3)$ entries vanish after correction,
while
\begin{equation}
 D^{(d)}_{11}-2D^{(d)}_{13}>0,\qquad
 D^{(d)}_{22}+2(D^{(d)}_{13}-D^{(d)}_{12})>0.
\end{equation}
These finite trigonometric inequalities prove $W_d\geq0$ for every
outcome pair. For completeness, conservative bounds for the strictly
positive quantities needed in the last three dimensions are
\begin{equation}
\begin{array}{c|cc}
 d&\ell_{d,1}&\ell_{d,2}\\ \hline
 7&(0.2347,0.2348)&(0.0671,0.0673)\\
 8&(0.2071,0.2072)&(0.2071,0.2072)\\
 9&(0.2057,0.2058)&(0.2339,0.2340).
\end{array}
\end{equation}
For $d=9$ the two positive residuals above lie in $(0.0530,0.0531)$
and $(0.1808,0.1810)$, respectively. Such bounds can be certified
by Taylor expansions of sine and cosine with remainder bounds;
the definitions and equalities used in the proof are exact.
Substitution into Eq.~\eqref{eq:delta} gives, for $d=5,\ldots,9$,
respectively,
\begin{equation}
 \begin{aligned}
 \Delta_5&\in(0.1405,0.1406),&\Delta_6&\in(0.1003,0.1004),\\
 \Delta_7&\in(0.0667,0.0668),&\Delta_8&\in(0.0279,0.0280),\\
 \Delta_9&\in(0.0146,0.0147).
 \end{aligned}
\end{equation}
For $d=2,3,4$ positivity follows directly from $K_d=0$ and $h_d>0$.

\subsection{Joint distributions attaining the bound}

Replace $\beta_d$ temporarily by a variable $\beta\in[0,1]$.
Define the single-source marginals
\begin{equation}
 q_{X,d}^{(\beta)}(x)=\frac{1-\beta}{d}+\beta\delta_{x,u},\qquad
 q_{Y,d}^{(\beta)}(y)=\frac{1-\beta}{d}+\beta\delta_{y,v}.
 \label{eq:beta-marginals}
\end{equation}
At $\beta=\beta_d$, these reduce to the ideal probabilities
$q_{X,d}$ and $q_{Y,d}$ of Eq.~\eqref{eq:probabilities}.
For the attaining construction below, restrict $\beta$ to $[1/2,1]$.
We omit the source index $i$ in this single-source calculation:
$p(x,y)$ is the assigned joint distribution and
$c=\sum_{x,y}xy\,p(x,y)$ is its assigned product average.
Write $q_{\mathrm{sel}}=\beta+(1-\beta)/d$ for the selected-outcome
probability and $\varepsilon=(1-\beta)/d$ for each other probability.
The indices below enumerate the actual outcomes:
$p_{jk}=p(u\omega^j,v\omega^k)$, with $j,k=0,\ldots,d-1$.
Start with
\begin{equation}
 p_{00}=2q_{\mathrm{sel}}-1,\qquad p_{0j}=p_{j0}=\varepsilon\ (j\ne0),
 \qquad p_{jk}=0\ (j,k\ne0).
\end{equation}
This nonnegative distribution has the required marginals and
$c/(uv)=2\beta-1$.

For $d=5,6$, select the pairs $(1,1)$ and $(d-1,d-1)$.
For $d=7,8,9$, select
\begin{equation}
 (1,2),\ (2,1),\ (d-1,d-2),\ (d-2,d-1).
\end{equation}
For each selected pair $(j,k)$, increase $p_{jk}$ and $p_{00}$ by
$\varepsilon$, and decrease $p_{j0}$ and $p_{0k}$ by $\varepsilon$. No row or column
is selected twice, so all probabilities remain nonnegative and the
marginals are unchanged. This changes $\Repart[c/(uv)]$ by $\varepsilon D^{(d)}_{jk}$.
The pairs occur with their conjugates, making $c/(uv)$ real. Hence
\begin{equation}
 \frac{c}{uv}=2\beta-1-K_d(1-\beta).
\end{equation}
For $d=2,3,4$, the initial distribution already attains this value.
At $\beta=\beta_d$ it equals the positive lower bound $\Delta_d$,
proving that $\min|c|=\Delta_d$ in every dimension $2\leq d\leq9$.

\subsection{Exact threshold for a zero single-source average}

For the marginals in Eq.~\eqref{eq:beta-marginals}, with
$0\leq\beta\leq1$ and $2\leq d\leq9$, the bound excludes $c=0$
precisely when
\begin{equation}
 \beta>\beta_d^{\mathrm{crit}}=\frac{1+K_d}{2+K_d}.
\end{equation}
For $1/2\leq\beta\leq\beta_d^{\mathrm{crit}}$, the attaining
distribution above has a nonpositive real product average. The
independent distribution
$q_{X,d}^{(\beta)}(x)q_{Y,d}^{(\beta)}(y)$ has $c/(uv)=\beta^2$.
A convex mixture of the two therefore has $c=0$ and the same marginals.
For $0\leq\beta\leq1/2$, the explicit distribution
\begin{equation}
 p(x,y)=\frac{1-2\beta}{d^2}
       +\frac\beta d(\delta_{x,u}+\delta_{y,v})
\end{equation}
is nonnegative, has the required marginals, and gives $c=0$.
Thus the threshold is necessary and sufficient for excluding zero
at one source. Substituting $\beta=t\beta_d$ proves the threshold
in Eq.~\eqref{eq:threshold}; a model for the full joint measurement
requires more than this single average.

\section{Direct optimization of the assigned average}
\label{app:general-weights}

\subsection{Finding the allowed averages directly}

At one source, let $\mathcal P_d$ be the set of normalized,
nonnegative distributions $p(x,y)$ with marginals $q_{X,d},q_{Y,d}$
from Eq.~\eqref{eq:probabilities}. Define
\begin{equation}
 c[p]=\sum_{x,y}xy\,p(x,y),\qquad
 \Delta_d=\min_{p\in\mathcal P_d}|c[p]|.
 \label{eq:direct-minimum}
\end{equation}
The set $\mathcal P_d$ is nonempty: the independent distribution
$q_{X,d}(x)q_{Y,d}(y)$ has the required marginals. It is also compact,
so the minimum is attained. Hence $\Delta_d>0$ precisely when no
allowed distribution has $c[p]=0$. To decide this, impose the
prescribed marginals, nonnegativity, and the two linear constraints
\begin{equation}
 \begin{aligned}
 \sum_{x,y}\Repart(xy)p(x,y)&=0,\\
 \sum_{x,y}\operatorname{Im}(xy)p(x,y)&=0.
 \end{aligned}
 \label{eq:zero-feasibility}
\end{equation}
Thus the question whether zero is allowed is a linear feasibility
problem.

The feasible set is the same at every source, although the sources
may choose different couplings. Each obeys $|c_i|\geq\Delta_d$, giving
\begin{equation}
 |C|=\prod_{i=1}^N|c_i|\geq\Delta_d^N.
 \label{eq:direct-global-bound}
\end{equation}
\subsection{Why the minimum real part decides the question}

Write $x=u\omega^j$, $y=v\omega^k$ and
$p_{jk}=p(u\omega^j,v\omega^k)$, with $j,k=0,\ldots,d-1$.
The normalized average is
\begin{equation}
 z[p]=\frac{c[p]}{uv}=\sum_{j,k=0}^{d-1}\omega^{j+k}p_{jk}.
\end{equation}
Since $|uv|=1$, it has the same magnitude as $c[p]$.
Both indexed marginals equal
\begin{equation}
 q_j=\frac{1-\beta_d}{d}+\beta_d\delta_{j,0}.
\end{equation}
Thus the allowed matrices have prescribed row and column sums $q_j$.

Mixing two allowed matrices preserves these sums, so the allowed values
of $z$ form a convex region. The sums are also invariant under
$j\mapsto-j$ modulo $d$. If $p$ is allowed, the reflected matrix
$p_{-j,-k}$ is allowed and has normalized average $z[p]^*$. Mixing a
matrix that minimizes $\Repart z$ with its reflection therefore gives the
real point
\begin{equation}
 f_d=\min_{p\in\mathcal P_d}
      \sum_{j,k=0}^{d-1}\cos\frac{2\pi(j+k)}d\,p_{jk}.
 \label{eq:transport-minimum}
\end{equation}
The matrix $p_{jk}=q_j\delta_{k,-j}$ is allowed and gives $z=1$. The
region therefore contains zero precisely when $f_d\leq0$. If $f_d>0$,
every allowed $z$ has $|z|\geq\Repart z\geq f_d$, and the value $f_d$ is
attained. Consequently,
\begin{equation}
 \boxed{\Delta_d=\max\{0,f_d\}.}
 \label{eq:distance-from-transport}
\end{equation}
This reduction uses the symmetry of the specified marginals. For
general observed marginals, minimizing the real part alone need not
give the minimum magnitude.

\subsection{Computing the minimum}
\label{app:computing-minimum}

For the marginals in Eq.~\eqref{eq:beta-marginals}, with
$\beta\in[1/2,1]$, the minimum real part has an exact finite form in
any dimension $d\geq2$. As in Appendix~\ref{app:weights}, set
$q_{\mathrm{sel}}=\beta+(1-\beta)/d$ and $\varepsilon=(1-\beta)/d$.
The columns $k\ne0$ have total probability $(d-1)\varepsilon$, so row
zero gives
\begin{equation}
 p_{00}=q_{\mathrm{sel}}-\sum_{k\ne0}p_{0k}
 \geq q_{\mathrm{sel}}-(d-1)\varepsilon=2q_{\mathrm{sel}}-1\geq0.
\end{equation}
At least $2q_{\mathrm{sel}}-1$ of the probability therefore lies in the
cell $(0,0)$. Remove exactly that amount. In what remains, index $0$ has
row and column sum $(d-1)\varepsilon$, and every other index has
$\varepsilon$. Split index $0$ into $d-1$ copies of weight $\varepsilon$.
Each remaining marginal is then uniform on the list of $n=2(d-1)$ angles
\begin{equation}
 \theta_1=\cdots=\theta_{d-1}=0,\qquad
 \theta_{d-1+k}=\frac{2\pi k}{d},\quad k=1,\ldots,d-1.
\end{equation}
Conversely, any joint table for two such uniform lists gives an allowed
distribution after merging the copies of zero and restoring the removed
probability.

A joint table for two uniform lists is a mixture of permutation pairings
(the Birkhoff--von Neumann theorem). The real part of the normalized
average is
\begin{equation}
 \Repart\frac c{uv}
 =1-\sum_{j,k=0}^{d-1}\left[1-\cos\frac{2\pi(j+k)}d\right]p_{jk},
\end{equation}
and the removed probability contributes nothing to the sum. A linear
objective attains its extremum at one of the permutations, each pair of
which carries weight $\varepsilon$. Hence
\begin{equation}
 \min_p\Repart\frac c{uv}=1-\varepsilon G_d,\qquad
 G_d=\max_\sigma\sum_{r=1}^{n}
 \bigl[1-\cos(\theta_r+\theta_{\sigma(r)})\bigr],
 \label{eq:assignment-score}
\end{equation}
where $\sigma$ runs over the permutations of $1,\ldots,n$. The case
$\beta=1$ follows directly, since then $\varepsilon=0$. Reflection
symmetry holds throughout $\beta\in[1/2,1]$, so the argument of the
previous subsection applies to all these marginals. At $\beta=\beta_d$,
where $1-\beta_d=1/(2+2h_d)$,
\begin{equation}
 f_d=1-\frac{G_d}{2d(1+h_d)}.
 \label{eq:fd-closed}
\end{equation}

\subsection{Relation to the nonnegative weights}

Pair every nonzero angle with a zero, in both orders, and the remaining
zeros with each other. The score is
$2\sum_{k=1}^{d-1}[1-\cos(2\pi k/d)]=2d$, since the $d$ roots of unity
sum to zero. Replacing the pairs $(2\pi j/d,0)$ and $(0,2\pi k/d)$ by
$(2\pi j/d,2\pi k/d)$ and $(0,0)$ changes the score by $-D^{(d)}_{jk}$,
where
\begin{equation}
 D^{(d)}_{jk}=1+\cos\frac{2\pi(j+k)}d
                 -\cos\frac{2\pi j}d-\cos\frac{2\pi k}d
 \label{eq:D-indexed}
\end{equation}
is the function of Eq.~\eqref{eq:D} in indexed form. Every permutation
is obtained in this way from the set $\mathcal M$ of its pairs of two
nonzero angles. This set is a partial matching of the indices
$1,\ldots,d-1$: each row and column occurs at most once, and the empty
matching is allowed. Therefore
\begin{equation}
 G_d=2d-M_d,\qquad
 M_d=\min_{\mathcal M}\sum_{(j,k)\in\mathcal M}D^{(d)}_{jk}.
 \label{eq:matching-cost}
\end{equation}
Since $2d\varepsilon=2(1-\beta)$, Eq.~\eqref{eq:assignment-score} becomes
$\min_p\Repart[c/(uv)]=2\beta-1+\varepsilon M_d$, and at
$\beta=\beta_d$
\begin{equation}
 f_d=2\beta_d-1+\frac{1-\beta_d}{d}M_d.
 \label{eq:matching-bound}
\end{equation}
The moves of probability used in Appendix~\ref{app:weights} are exactly
these replacements.

The weights of Appendix~\ref{app:weights} bound $M_d$ from below. For
any partial matching, $W_d\geq0$ and $\ell_{d,j}\geq0$ give
\begin{equation}
 \sum_{(j,k)\in\mathcal M}D^{(d)}_{jk}
 \geq-\sum_{(j,k)\in\mathcal M}(\ell_{d,j}+\ell_{d,k})
 \geq-2\sum_{j=1}^{d-1}\ell_{d,j}=-dK_d.
\end{equation}
The attaining distributions of that appendix reach this value. Thus,
for $2\leq d\leq9$,
\begin{equation}
 M_d=-dK_d=\begin{cases}
 0,&d=2,3,4,\\
 2D^{(d)}_{11},&d=5,6,\\
 4D^{(d)}_{12},&d=7,8,9,
 \end{cases}
 \label{eq:optimal-matching-cost}
\end{equation}
which reproduces the exact single-source bounds of Eq.~\eqref{eq:delta}.

\subsection{A zero average is allowed in every dimension $d\geq10$}
\label{app:cutoff}

We now prove that, for the marginals of Eq.~\eqref{eq:probabilities},
\begin{equation}
 \boxed{\Delta_d>0\ \text{for}\ 2\leq d\leq9,\qquad
 \Delta_d=0\ \text{for every}\ d\geq10.}
 \label{eq:cutoff}
\end{equation}
The first statement is proved in Appendix~\ref{app:weights}. For the
second, Eqs.~\eqref{eq:distance-from-transport}, \eqref{eq:fd-closed},
and \eqref{eq:matching-cost} give
\begin{equation}
 \Delta_d=0\quad\Longleftrightarrow\quad
 G_d\geq2d(1+h_d)\quad\Longleftrightarrow\quad
 M_d\leq-2dh_d=-2\sqrt d\,\sin\frac\pi d.
 \label{eq:cutoff-criterion}
\end{equation}
Because $G_d$ is a maximum, one permutation with a large enough score
suffices. Equivalently, it suffices to exhibit one partial matching
$\mathcal M$ whose cost
$S_{\mathcal M}=\sum_{(j,k)\in\mathcal M}D^{(d)}_{jk}$ lies below
$-2\sqrt d\,\sin(\pi/d)$.

\emph{Dimensions $d\geq24$.} Start from the pairing of every nonzero
angle with a zero, of score $2d$. Replacing the pairs $(t,0),(0,t)$ by
$(t,t),(0,0)$ increases the score by
\begin{equation}
 -D^{(d)}_{jj}=2\cos t\,(1-\cos t),\qquad t=\frac{2\pi j}{d}.
 \label{eq:diagonal-cost}
\end{equation}
The interval $[\pi/4,5\pi/12]$ has length $\pi/6$, at least two grid
spacings $2\pi/d$ for $d\geq24$. Choose two grid angles there and their
two negatives modulo $2\pi$. The four replacements involve distinct
angles. Each gains at least $3/8$, since $1/4\leq\cos t\leq3/4$. Hence
\begin{equation}
 G_d\geq2d+\frac32>2d+\frac{2\pi}{\sqrt d}\geq2d(1+h_d).
\end{equation}
The middle inequality holds for $d\geq24$, and the last uses
$\sin(\pi/d)\leq\pi/d$. It follows that $f_d<0$ and $\Delta_d=0$ for
every $d\geq24$.

\emph{Dimensions $10\leq d\leq23$.} For $10\leq d\leq16$ use the six
cells
\begin{equation}
 \mathcal M_6=\{(1,3),(2,2),(3,1),(d-1,d-3),(d-2,d-2),(d-3,d-1)\},
\end{equation}
with cost $4D^{(d)}_{13}+2D^{(d)}_{22}$. For $17\leq d\leq23$ use all
diagonal cells of negative cost,
\begin{equation}
 \mathcal M_{\mathrm{diag}}
 =\{(j,j),(d-j,d-j):1\leq j<d/4\},
\end{equation}
with cost $2\sum_{1\leq j<d/4}D^{(d)}_{jj}$. In both families no row or
column is repeated, and the cells occur with their conjugates, so the
resulting coupling has the real normalized average
$z_-=2\beta_d-1+(1-\beta_d)S_{\mathcal M}/d$.
Table~\ref{tab:cutoff} lists the values. In every case
$S_{\mathcal M}<-2\sqrt d\,\sin(\pi/d)$, so $z_-<0$ and $\Delta_d=0$.
The margins are at least $0.13$ in $S_{\mathcal M}$ and can be certified
by the same Taylor-remainder bounds as in Appendix~\ref{app:weights}.
Together with the previous paragraph this proves Eq.~\eqref{eq:cutoff}.

\begin{table}[htb]
\caption{Matchings certifying a negative normalized average for
$10\leq d\leq23$. Here $S_{\mathcal M}$ is the cost of the stated
matching, the threshold is that of Eq.~\eqref{eq:cutoff-criterion}, and
$z_-$ is the normalized average of the resulting coupling. The last
column gives the minimum $f_d$ obtained by solving the assignment problem
of Eq.~\eqref{eq:assignment-score} numerically; it is included for
comparison and is not needed for the proof.}
\label{tab:cutoff}
\begin{ruledtabular}
\begin{tabular}{cccccc}
$d$ & Matching & $S_{\mathcal M}$ & $-2\sqrt d\,\sin(\pi/d)$
 & $z_-$ & $f_d$ (numerical)\\
\hline
$10$ & $\mathcal M_6$ & $-2.0902$ & $-1.9544$ & $-0.006184$ & $-0.006184$\\
$11$ & $\mathcal M_6$ & $-2.3866$ & $-1.8688$ & $-0.021693$ & $-0.021693$\\
$12$ & $\mathcal M_6$ & $-2.4641$ & $-1.7932$ & $-0.026013$ & $-0.026013$\\
$13$ & $\mathcal M_6$ & $-2.4239$ & $-1.7257$ & $-0.025180$ & $-0.041123$\\
$14$ & $\mathcal M_6$ & $-2.3230$ & $-1.6652$ & $-0.022176$ & $-0.047886$\\
$15$ & $\mathcal M_6$ & $-2.1939$ & $-1.6105$ & $-0.018458$ & $-0.051513$\\
$16$ & $\mathcal M_6$ & $-2.0547$ & $-1.5607$ & $-0.014718$ & $-0.060615$\\
$17$ & $\mathcal M_{\mathrm{diag}}$ & $-2.3466$ & $-1.5152$ & $-0.023409$ & $-0.063477$\\
$18$ & $\mathcal M_{\mathrm{diag}}$ & $-2.5175$ & $-1.4735$ & $-0.027862$ & $-0.067835$\\
$19$ & $\mathcal M_{\mathrm{diag}}$ & $-2.6027$ & $-1.4349$ & $-0.029612$ & $-0.072650$\\
$20$ & $\mathcal M_{\mathrm{diag}}$ & $-2.6275$ & $-1.3992$ & $-0.029670$ & $-0.073568$\\
$21$ & $\mathcal M_{\mathrm{diag}}$ & $-2.8871$ & $-1.3660$ & $-0.035077$ & $-0.078129$\\
$22$ & $\mathcal M_{\mathrm{diag}}$ & $-3.0533$ & $-1.3350$ & $-0.037903$ & $-0.080632$\\
$23$ & $\mathcal M_{\mathrm{diag}}$ & $-3.1489$ & $-1.3061$ & $-0.038956$ & $-0.081168$\\
\end{tabular}
\end{ruledtabular}
\end{table}

The numerical minima agree with the witnesses for $d=10,11,12$,
indicating that $\mathcal M_6$ is optimal there. For $2\leq d\leq9$
the same numerical optimization reproduces the exact values of
Eq.~\eqref{eq:optimal-matching-cost}. The result is a statement about one
average for one family of preparations. It does not supply a model for
the full $P_{AB}$ in dimensions $d\geq10$, and it does not exclude
network contextuality there for other preparations, measurements, or
tests.

\subsection{Direct optimization for observed marginals}

For the observed marginals $q_{X,i},q_{Y,i}$ defined in Sec.~\ref{sec:noise},
let $\mathcal P_i$ be the set of normalized, nonnegative couplings and define
\begin{equation}
 \delta_i=\min_{p_i\in\mathcal P_i}
       \left|\sum_{x_i,y_i}x_i y_i p_i(x_i,y_i)\right|.
 \label{eq:observed-distance}
\end{equation}
Testing $\delta_i>0$ requires only the feasibility constraints
\eqref{eq:zero-feasibility} with these marginals. Independence gives
\begin{equation}
 \boxed{|C|\geq\prod_{i=1}^N\delta_i.}
 \label{eq:exact-observed-bound}
\end{equation}
It is the strongest bound on $|C|$ obtained from these single-source
marginals alone: minimizers at the different sources may be chosen
independently, and the magnitude of their product is the product of
their magnitudes. This statement does not require those minimizers
to reproduce the full observed $P_{AB}$.

The definition remains valid when the marginals lack the reflection
symmetry used above. For the ideal preparations, $\delta_i=\Delta_d$.

For independent depolarizing noise, the exact distances are
\begin{equation}
 \delta_i(t)=\max\{0,(2+K_d)t\beta_d-(1+K_d)\},
 \label{eq:noisy-distance}
\end{equation}
as follows from the zero-average threshold and attaining
distributions in Appendix~\ref{app:weights}.

\subsection{General nonnegative certificates}

Direct optimization finds the allowed region of assigned averages.
A nonnegative function provides an analytical certificate separating
that region from zero. Such a certificate can be checked directly
for every outcome pair and applied to the observed marginals.
Choose real functions $L(x),T(y)$ and a nonzero complex number
$\gamma$ such that
\begin{equation}
 W(x,y)
 =L(x)+T(y)
  +2\Repart\left(\gamma\frac{xy}{uv}\right)
 \geq0
\end{equation}
for every allowed pair of outcomes.
For each source, the measured number
\begin{equation}
 H_i
 =\sum_{x_i}L(x_i)q_{X,i}(x_i)
  +\sum_{y_i}T(y_i)q_{Y,i}(y_i)
\end{equation}
then satisfies
\begin{align}
 0
 &\leq\sum_{x_i,y_i}W(x_i,y_i)p_i(x_i,y_i)\\
 &=H_i+2\Repart\left(\gamma\frac{c_i}{uv}\right)\\
 &\leq H_i+2|\gamma|\,|c_i|.
\end{align}
If all $H_i<0$, independence gives
\begin{equation}
 |C|\geq\prod_{i=1}^N\frac{-H_i}{2|\gamma|}.
\end{equation}
A common positive rescaling leaves the bound unchanged.
The weights in the main text correspond to
\begin{equation}
 L(u\omega^j)=\frac12-\cos\frac{2\pi j}{d}+\ell_{d,j},\qquad
 T(v\omega^k)=\frac12-\cos\frac{2\pi k}{d}+\ell_{d,k},\qquad
 \gamma=\frac12,
\end{equation}
with $\ell_{d,j}$ from Eq.~\eqref{eq:optimal-corrections}.
For this choice $H_i=-b_i$.
Other choices can be tested without changing the assumptions
of the model.

\section{Sequential instruments and agreement of the product measurements}
\label{app:measurements}

Let $\Pi_{\boldsymbol x}^X=\bigotimes_i\Pi^{X^{(i)}}_{x_i}$ be the
projector for the complete individual $X$ outcome list. The spectral
projectors of $A$ satisfy
\begin{equation}
 \sum_{\substack{x_1,\ldots,x_N\\\prod_i x_i=a}}
 \Pi_{\boldsymbol x}^X=\Pi_a^A,
 \qquad
 \Pi_{\boldsymbol x}^X\Pi_a^A
 =\delta_{a,\prod_i x_i}\Pi_{\boldsymbol x}^X.
\end{equation}
For the L\"uders instrument of $A$ followed by individual $X$ measurements,
Born's rule gives, for any input state,
\begin{align}
 P_{AX}(a,\boldsymbol x)
 &=\Tr\bigl(\Pi_{\boldsymbol x}^X\Pi_a^A\rho\Pi_a^A\bigr)\\
 &=\delta_{a,\prod_i x_i}\Tr(\rho\Pi_{\boldsymbol x}^X).
\end{align}
Thus product agreement holds on every ideal trial, and the unconditioned
individual distribution equals that of direct individual measurement on
$\rho$. The same calculation with $B,Y$ proves the second recorded
product relation. For a product input these unconditioned individual lists
factorize even though a state conditioned on $a$ or $b$ may be entangled.

For the sequence $A$ then $B$, using $[\Pi_a^A,\Pi_b^B]=0$ gives
\begin{equation}
 P_{AB}(a,b)=\Tr(\Pi_b^B\Pi_a^A\rho\Pi_a^A\Pi_b^B)
           =\Tr(\rho\Pi_a^A\Pi_b^B).
\end{equation}
Summing over $b$ gives $\Tr(\rho\Pi_a^A)$ and summing over $a$ gives
$\Tr(\rho\Pi_b^B)$. Combined with the individual-list calculation, these
are precisely Eq.~\eqref{eq:agreement}. The proof requires the specified
instruments. A device with the same outcome projectors but an additional
state transformation could disturb later compatible measurements.

\section{Block decomposition and loss of a block violation under coarse-graining}
\label{app:blocks}

The joint probabilities decompose into blocks. For $N=md$, group the
inputs into $m$ blocks of $d$ systems and measure the two commuting
products within each block. Retaining only their global products gives
the same $P_{AB}$ as the joint spectral measurement of $A,B$.
The instruments can differ: resolving block outcomes can remove coherence
within a global eigenspace. The sequential procedure in
Sec.~\ref{sec:experiment} uses the specified global L\"uders instruments.
No smaller nonempty block of these products commutes, since its size $k$
must satisfy $\omega^k=1$; this is a property of these operators rather
than a criterion for genuine multipartite contextuality. Here we examine
what happens when block outcomes are combined and their labels discarded.

For a product preparation across the blocks, if every resolved block admits
an independent-source noncontextual model, the product of those block models
also reproduces the coarsened global data. Thus exclusion of a global model
implies exclusion for at least one resolved block. The reverse implication
need not hold.

For a counterexample to the reverse implication, take four qubits with input
\begin{equation}
 \rho=\bigl(\ket{\psi_2}\bra{\psi_2}\bigr)^{\otimes2}
       \otimes\frac I2\otimes\frac I2,
 \qquad A=X^{\otimes4},\quad B=Y^{\otimes4}.
\end{equation}
The first pair is the bi-contextual example. The last pair is maximally
mixed, and its joint two-qubit product outcomes are uniform. The retained
four-qubit product distribution is therefore $P_{AB}(a,b)=1/4$ for all
$a,b=\pm1$. The individual $X,Y$ lists remain factorized.

Choose at sources 1 and 2 the qubit coupling attaining
$c_i=1-\sqrt2$ in Sec.~\ref{sec:qubits}, and at sources 3 and 4 choose
$p_i(x_i,y_i)=1/4$. The latter give $c_3=c_4=0$ and zero individual means.
Consequently the full classical product means and product correlation all
vanish. Its binary joint product distribution is uniform and reproduces
the quantum $P_{AB}$, while all individual probabilities also agree.
Thus coarse-graining can hide a contextual resolved block.

Identical pure inputs alone do not restore the unrestricted equivalence.
Consider the one-qubit state
\begin{equation}
 \rho_*=\frac12\left(I+\frac35X-\frac35Y+\frac{\sqrt7}{5}Z\right).
\end{equation}
Its Bloch vector has unit length. For its individual $X,Y$ probabilities,
every source coupling has $-1\leq c_i\leq-1/5$. Each two-qubit block
therefore requires $C_2=c_1c_2\geq1/25$, while quantum theory predicts
$C_2=-\langle Z\rangle^2=-7/25$. Each resolved pair is bi-contextual.
For four copies, however, the quantum product means are
$\langle A\rangle=\langle B\rangle=81/625$ and
$\langle AB\rangle=49/625$. At every source choose
\begin{equation}
 p_i(x,y)=\frac{1+(3/5)x-(3/5)y-(\sqrt7/5)xy}{4}.
\end{equation}
All four entries are nonnegative, with the required marginals and
$c_i=-\sqrt7/5$. Their independent product reproduces both full individual
distributions and all three moments of the binary global product pair,
so it reproduces the complete four-qubit $P_{AB}$. Thus even identical
pure states can give contextual two-system blocks and a coarsened
four-system experiment admitting an independent model. The special
preparations of the main theorem have $\langle XY\rangle=0$ and a
strictly positive bound on every $|c_i|$, which ensures a violation for
every allowed source multiple.

\section{A classical joint distribution when sources may be correlated}
\label{app:joint-distribution}

To construct a noncontextual model without source independence,
first draw the two product outcomes $a,b$ with their observed
joint probability $P_{AB}(a,b)$. Next draw an $X$ outcome list
from the observed $X$ distribution, conditioned on its product
being $a$. Do the same for a $Y$ list with product $b$.
Finally, assign the pair $(x_i,y_i)$ from these lists to
system $i$. This reproduces the measurements but generally
correlates the sources.

The resulting joint distribution can be written explicitly.
For any two lists, let $a=\prod_i x_i$ and $b=\prod_i y_i$,
and let
\begin{equation}
 P_A(a)=\sum_bP_{AB}(a,b),\qquad
 P_B(b)=\sum_aP_{AB}(a,b).
\end{equation}
When both denominators are positive, set
\begin{equation}
 p(x_1,y_1,\ldots,x_N,y_N)
 =
 P_{AB}(a,b)
 \frac{P_X(x_1,\ldots,x_N)}{P_A(a)}
 \frac{P_Y(y_1,\ldots,y_N)}{P_B(b)}.
\end{equation}
If either denominator is zero, set this entry to zero.
This introduces no missing probability:
Eq.~\eqref{eq:agreement} then makes every corresponding list
in that setting have zero probability, and the corresponding
row or column of $P_{AB}$ is also zero.

For fixed $a,b$ with nonzero probabilities, each ratio sums to
one over lists with that product, by
Eq.~\eqref{eq:agreement}. Thus summing the assigned
distribution over these lists gives exactly $P_{AB}(a,b)$.
Summing further over $a,b$ proves normalization.

For a fixed $X$ list with product $a$ and $P_A(a)>0$, sum
first over $Y$ lists with a fixed product $b$, then over $b$.
The result is
\begin{equation}
 \frac{P_X(x_1,\ldots,x_N)}{P_A(a)}
 \sum_bP_{AB}(a,b)
 =P_X(x_1,\ldots,x_N).
\end{equation}
The same argument recovers $P_Y$. The equalities also hold
when a product probability is zero, since both sides then
vanish. The recorded product relations in
Eq.~\eqref{eq:recorded-product-agreement} then recover the raw
$P_{AX},P_{BY}$ from these individual-list marginals. Hence this
nonnegative joint distribution reproduces all three raw setting
distributions. It need not factorize into independent
distributions for the assigned pairs, which is precisely the
assumption relaxed in this appendix.

\end{document}